\documentclass[aps,prb,twocolumn,floatfix,showpacs,superscriptaddress]{revtex4-2}
\usepackage{float}
\usepackage{epsfig}
\usepackage{times}
\usepackage{bm}
\usepackage{braket}
\usepackage{color}
\usepackage{hyperref}

\newcommand{\e}{{\rm e}}

\usepackage{amsmath,amssymb}	

\begin{document}
\title{Exotic coupled spin-charge states in decorated honeycomb magnets: A hybrid-Monte Carlo study}

\author{Satyabrata Jana}
\email{satyabrataj.physics.rs@jadavpuruniversity.in}
\affiliation{Department of Physics, Jadavpur University, Kolkata 700032, West Bengal, India}

\author{Sahinur Reja}
\email{sahinurreja.physics@jadavpuruniversity.in}
\affiliation{Department of Physics, Jadavpur University, Kolkata 700032, West Bengal, India}

\begin{abstract}
We uncover four exotic coupled spin-charge ground states in the strong coupling limit of the Kondo lattice model at various electronic fillings on a frustrated decorated honeycomb lattice, where each regular honeycomb sublattice point is occupied by three-site triangular units. We employ a hybrid Markov Chain Monte Carlo (hMCMC) simulation method which combines classical MCMC for localized spins and exact diagonalization of the electronic Hamiltonian. 
Two of the spin-charge ground states, respectively consists of three-site and six-site ferromagnetic (FM) clusters arranged in anti-FM and $120^{\circ}$ Yafet-Kittel (YK) phase which we label as S-AF (super-antiferromagnet) and S-YK (super-YK) respectively. Two even more interesting coupled spin-charge states, respectively accommodate FM dimers and trimers (as three-site line segment), which we label as FM-D and FM-T. In both cases, the anti-FM aligned dimers and trimers in respective phases, are arranged in stripes along one of three lattice directions: the spontaneously symmetry broken phases giving rise to non-trivial macroscopic degeneracy. These underlying magnetic textures (except S-YK state) restrict electrons in fragmented small regions (e.g, triangular units, two-site dimers, three-site line segments respectively in S-AF, FM-D and FM-T), resulting in flat bands by opening large gaps in electronic density of states, which in turn stabilize these coupled spin-charge states: a "band effect". These exotic spin-charge ground states could be relevant to electron-doped spin-systems resulting from various metal-organic frameworks (MOFs), which have attracted significant attention to condensed matter physics. 
\end{abstract}

\maketitle

\section{introduction}
Over the decades, it has been understood that the magnetic frustration arising from lattice geometry and/or magnetic interactions plays a crucial role in determining the nature of magnetic ground states. The simplest case of magnetic frustration arises when we try to arrange anti-ferromagnetically (AFM) coupled spins on a triangular lattice. The frustration in spin orientation gives rise to a large number of degenerate spin configurations that forces the system to be in exotic magnetic state, such as spin glass,\cite{Spin_Glass_S_F_Edwards_1975, Spin_Glass_RevModPhys.58.801} spin ice,\cite{Spin_Ice_Bramwell_2002,Spin_Ice_Castelnovo_2008} spin liquids \cite{Spin_Liquid_RVB_ANDERSON1973153,Frustration_Spin_liquid_Balents_2010}. So, magnetically frustrated systems are a fertile playground to explore quantum spin liquids in particular due to their intrinsic connection to quantum computing \cite{Frustration_Spin_liquid_Balents_2010}. 

It is well known that the simple magnetic interactions (FM, AFM etc) in various lattice systems often give rise to rather mundane collinear and coplanar magnetic textures. However, including conduction electrons in these magnetic systems would drastically modify the nature of the magnetic ground state. In general, the impact of conduction electrons on magnetic structures on frustrated lattices has attracted huge attention to magnetic materials research community. These systems with localized spins plus conduction electrons modelled by the so-called Kondo Lattice model (KLM), lead to the conceptual development in magnetism and electron transport, such as Ruderman-Kittel-Kasuya-Yosida (RKKY) interaction, Kondo effect and Double-exchange mechanism \cite{RKKY_Interaction_Ruderman_1954, Ferro_AF_Exchange_Interaction_10.1143/PTP.16.45, RK_Zener_Yosida_1957, Kondo_10.1143/PTP.32.37, Double_Exchange_Zener_1951, Double_Exchange_PhysRev.118.141}

Unlike the simple spin model (e.g, Heisenberg model), this model on frustrated lattice geometries, e.g triangular lattice, can support coupled spin-charge order (with non-collinear magnetic spin arrangement)\cite{Non_Colinear_PhysRevB.91.140403}
or even non-coplanar spin textures like Magnetic Skyrmion: a topologically protected local whirl of the spin configuration \cite{Skyrmion_Science,Skyrmion_Nature,Skyrmion_Nagaosa2013}.
In this context, our recent study suggests that the coupled spin-charge systems can provide an electronic route to stabilize skyrmion-like non-coplanar magnetic textures\cite{Electronic_Route_PhysRevB.93.155115}.
Nanoscale-sized and topological stability properties make mobile magnetic skyrmions a potential candidate for dense magnetic data storage devices using race track memory setup\cite{Racetrack_Memory}.
Furthermore, conduction electrons are shown to have profound effects on the Skyrmion crystal phases\cite{Skyrmionic_Crystal_Phase_Jana_2023}
found in triangular lattice systems. Even more interestingly, these non-coplanar magnetic ground states often drive the conduction electrons to acquire Berry phase and to form topologically non-trivial electronic bands giving rise to Quantum Anomalous Hall effects \cite{Anomalous_Hall_effect_RevModPhys.82.1539,Quantum_Hall_Effect_PhysRevLett.101.156402,Double_Exchange_PhysRevLett.105.216405}.

In this study, we explore the exotic magnetic phases in a coupled spin-charge system on a decorated honeycomb lattice (DHL) where each honeycomb site is occupied by a 3-site triangular unit. We are interested in DHL as it plays a prominent role in different branches of physics. Most importantly, the DHL is known to host a chiral spin liquid as an exact ground state of the Kitaev model\cite{kivelson_chiral_spin_liquid_PhysRevLett.99.247203}. It plays an important role in ultracold
atoms\cite{ultracold_DHL_PhysRevA.90.053627} in the context of quantum magnets\cite{ultracold_DHL_PhysRevA.90.053627,quantm_magnets_dhl_ultracold_PhysRevB.81.134418} and topological insulators\cite{topo_insu_dhl_ultracold_PhysRevB.81.205115} as the DHL shows distinctive, remarkable features different from the kagome and honeycomb lattices. Also, strongly correlated electron systems modeled by the Hubbard model have been extensively studied on DHL as an simplified lattice out of metal-organic framework (MOF) $\mathrm{Mo_3S_7(dmit)_3}$ in the context of exotic insulating states and Haldane phase\cite{hubbard_dhl_ben_1_PhysRevB.104.075104,hubbard_dhl_ben_2_PhysRevB.94.214418,hubbard_dhl_ben_3_PhysRevB.103.L081114}. Furthermore, this lattice system exhibits the flat bands in electronic spectra which is shown to accommodate unconventional superconductivity \cite{PhysRevB.109.075153} at relatively high temperatures and also to host fractional Chern insulators or topological superconductivity \cite{PhysRevB.83.220503}.  

Using the hybrid-MCMC method, we simulate the strong coupling KLM at various electronic fillings on DHL. We found four exotic coupled spin-charge ground states in the intermediate AFM Heisenberg couplings. Two of the states consist of superstructures FM clusters arranged in AFM and $120^{\circ}$ Yafett-Kittel orders respectively. The other two magnetic textures are made of, respectively, FM dimers (two-site) and trimers (three-site line segment) that are anti-FM aligned. Interestingly, both magnetic textures possess macroscopic degeneracy as there are multiple ways to tile the DHL with dimers and trimers. 

The paper is organized as follows: we describe the Kondo lattice model we have considered on a decorated honeycomb lattice in section ({\bf II}). Next in section ({\bf III}), the hybrid-MCMC methods and the physical quantities (density of states, spin-structure factors, etc) used to analyze the ground state phases have been discussed. The numerical results deducing ground state phase diagrams for various fillings have been presented in section ({\bf IV}). Finally, we summarize the results in section  ({\bf V}).

\begin{figure}[htb]
	\centering
	\includegraphics[width=0.8\linewidth,trim={2.5cm 4.1cm 8.33cm 15.68cm},clip]{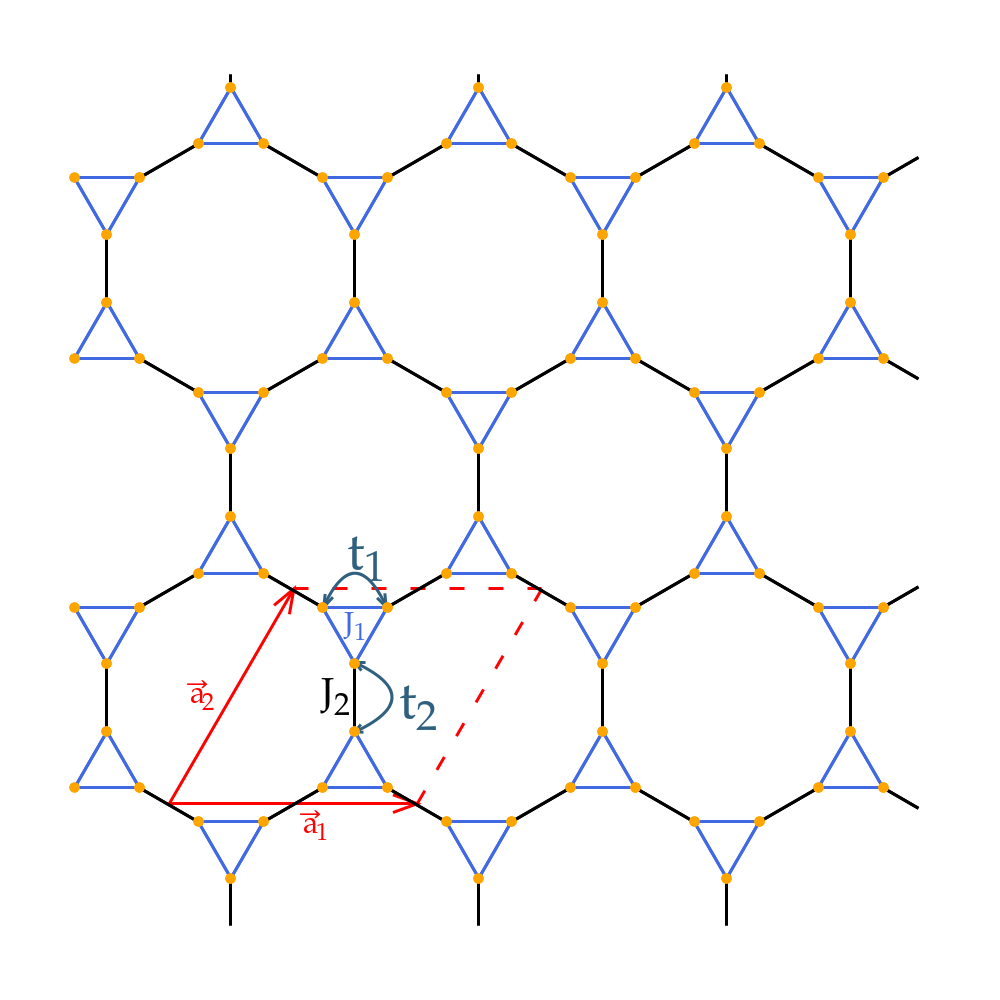}
	\caption{(Color online) Schematic of unit cell of decorated honeycomb lattice consisting of 6-sites (filled circle) defined by lattice vectors $\vec{a_1}$ and $\vec{a_2}$. The electron hopping (Heisenberg exchange) $t_1 (J_1)$ and $t_2 (J_2)$ are on intra-triangular (blue) and inter-triangular (black) bonds respectively.} 
	\label{fig:Unit_Cell}
\end{figure}

\section{Model and Method}
To study the exotic magnetic ground states, we start from the Kondo Lattice model (KLM) on a decorated honeycomb lattice (DHL) consisting of localized magnetic moments coupled to conduction electrons. As shown in Fig.\ref{fig:Unit_Cell}, the DHL we considered is described by a 6-site unit cell defined by lattice vectors $\vec{a_1}$ and $\vec{a_2}$. The intra-triangular and inter-triangular hopping strengths are given by $t_1$ and $t_2$. Assuming the magnetic moments to be classical, the model in the strong Kondo coupling limit reduces to so called Double Exchange (DE) model with additional antiferromagnetic coupling between the localized moments \cite{Double_Exchange_PhysRevLett.105.216405,Double_Exchange_DAGOTTO20011,Double_Exchange_PhysRevB.54.R6819,Double_Exchange_book}. The classical approximation of localized spins is a good starting point, unless the localized moments are spin-$1/2$\cite{Classical_Approximation_PhysRevB.58.6414,Classical_Approximation_PhysRevB.73.224405}.
The resulting Hamiltonian on DHL becomes:
\begin{equation} 
    \label{eq:effective Equation}
    \begin{split}
        H = & -\sum_{\langle i,j\rangle } \tilde{t}_1 (c_i^\dagger c_j + {h.c.})-\sum_{\langle i,j\rangle } \tilde{t}_2 (c_i^\dagger c_j + {h.c.})\\
        & + J_1 \sum_{\langle i,j\rangle} S_i \cdot S_j + J_2 \sum_{\langle i,j\rangle} S_i \cdot S_j
    \end{split}
\end{equation}
where $c_i(c_i^{\dagger})$ is the usual annihilation (creation) operator for a electron at the site $i$ with its spin parallel to the local magnetic moment $S_i$. The angular bracket $\langle i,j \rangle$ is used to represent the NN interacting pair of sites on DHL. The Heisenberg coupling strengths between NN localized spins on intra-triangular and inter-triangular bonds are represented, respectively by $J_1$ and $J_2$. Due to the strong Kondo coupling approximation, the electron spin at site $i$ has to point along the local quantization axis parallel to $S_i$. So, the effective electron hopping integrals becomes $\tilde{t}_{1,2}=t_{1,2}\left[\cos({\theta _i/2})\cos({\theta _j/2}) + \sin({\theta _i/2})\sin({\theta _j/2})\e^{-i(\phi_i -\phi_j)}\right]$ between site $i$ and $j$ which now depends on the orientation of localized spins at site $i$ and $j$ defined by polar and azimuthal angle $(\theta_i,\phi_i)$ and $(\theta_j,\phi_j)$ respectively. Further, $t_1$ and $t_2$ are the bare electron hopping strengths on intra-triangular and inter-triangular bonds on DHL. The parameters of the model are the bare electron hopping strengths $t_1$, $t_2$; AFM Heisenberg couplings $J_1$ and $J_2$; and electron filling $n_e$. In this study, we set the unit of energy as $t_1=t_2=1$ and explore the ground state phase diagram of the model system defined in Eq.\ref{eq:effective Equation} at different electronic fillings $n_e=1/3, 1/2$ and $2/3$ along the line defined by $J_1=J_2$. This choice of parameter space represents a maximally frustrated lattice. 

To simulate the model in Eq.(\ref{eq:effective Equation}), we employ the so-called hybrid Markov Chain Monte Carlo (hMCMC) method which includes the usual MCMC for classical localized spins and exact diagonalization of electronic degrees of freedom \cite{PhysRevLett.80.845}. As the electron hopping Hamiltonian (first two terms in Eq.(\ref{eq:effective Equation}) depends on the localized spin orientations, the solution of a electronic problem is needed at each Monte Carlo update of localized spins in order to obtain the electronic contribution to the total energy of a given classical configuration of localized spins. The electronic energy is calculated by filling the single particle states by electrons in the system defined by the electronic density $n_e$. Notice that the time-consuming diagonalization of electronic Hamiltonian does not allow to simulate larger sized systems as we are familiar with classical MCMC methods. We use simulated annealing i.e, the MCMC simulation starts at high temperature and slowly moves to low temperature in steps upto $T=0.001t_1$, i.e {\it stimulated annealing}. At each temperature step, we use about $10^4$ MCMC steps for equilibration and similar number of steps for measurement of ground state properties. Since our primary focus in this paper is the ground-state phase (phases at the lowest temperature possible) diagram rather than temperature-induced phenomena, we supplement the hMCMC procedure with a subsequent numerical optimization initialized from the final hMCMC configuration. Also, after getting idea from real space spin configuration of certain ground state from hMCMC simulation, we sometimes incorporate simple variational calculations to optimize the ground state energy (see below).

We have used $N=54 (=3\times 3\times 6$) site system with periodic boundary conditions at electronic filling $n_e=1/2, 1/3$ and $2/3$. This relatively smaller system allows us to diagonalize full $54\times 54$ hopping Hamiltonian for the calculation of electronic energy at each MCMC update for localized spin and captures all the exotic coupled spin-charge ground states to be presented in this paper. However, to substantiate our results, we also simulate $N=216 (=6\times 6\times 6$) site system with periodic boundary condition using a variant of hMCMC method called travelling cluster approximation (TCA)\cite{kumar2004travellingclusterapproximationstrong}. The basic idea relies on the approximation that an MCMC update for a localized spin at site $i$ modifies the hopping matrix elements only around that given site. So, this approximation allows us to calculate the electronic energy contribution by diagonalizing a smaller cluster of $N_c$ sites with $N_c\leq N$ built around site $i$ compatible with the lattice symmetry. The only thing we need to take care of is that the cluster now would accommodate $n_eN_c$ electrons to fill the eigenstates. 

We mostly identify various magnetic phases by visualizing the real-space configurations of the hMCMC simulations. However, for further explanations of the phases, we calculate the physical quantities: the electronic density of states given by,
$$D(\omega) \propto \frac{1}{N}\sum\delta( \omega- \epsilon_i)$$ where $\epsilon_i$ being discrete eigen energies obtained by diagonalizing the hopping Hamiltonian. The delta function is approximated by Lorentzian with a broadening factor. Also, to confirm the magnetic ordering present in a particular state we compute the localized spin structure factor at reciprocal lattice point $q$ using 
$$S(q) \propto \sum_{ij} \langle S_i\cdot S_j \rangle e^{-iq.(r_i-r_j)}$$
where the $\langle \cdots \rangle$ represents MCMC average, $r_i$ is position vectors of site $i$. We calculate the ground state electronic charge density at site $i$ ($n_i^0$) from tight-binding state obtained by diagonalizing the electronic Hamiltonian.

 \section{Numerical results}
As mentioned earlier, we are interested in the maximally frustrated version of the model in Eq.(\ref{eq:effective Equation}), leaving only one parameter $J_1=J_2$ with energy unit $t_1=t_2=1$ in the model. So, we ran the extensive hMCMC simulations at various electronic fillings $n_e$ with varying $J_1$, each starting at high temperature (T) and stepwise going down to very low $T=0.001t_1$. As we are interested in ground state phases rather than the temperature induced phenomena in this system, we finally extracted the sample of real space spin configurations and corresponding MCMC averages of physical properties at the lowest T, such as: total energy; electronic density of states $\rho (E)$ and charge density $n_i^0$; spin structure factors $S(q)$. In all the cases, it has been checked that the hMCMC simulations in a plane (say, $X-Y$ plane) always give lower energy than that of the simulations in 3-dimensional space.   

\subsection{ Phases at n$_\text{e}$ = 2/3}
\begin{figure}[hbt!]
	\centering
	\includegraphics[width=0.7\linewidth,trim={0.55cm 1.1cm 0.4cm 0.5cm},clip,angle=270]{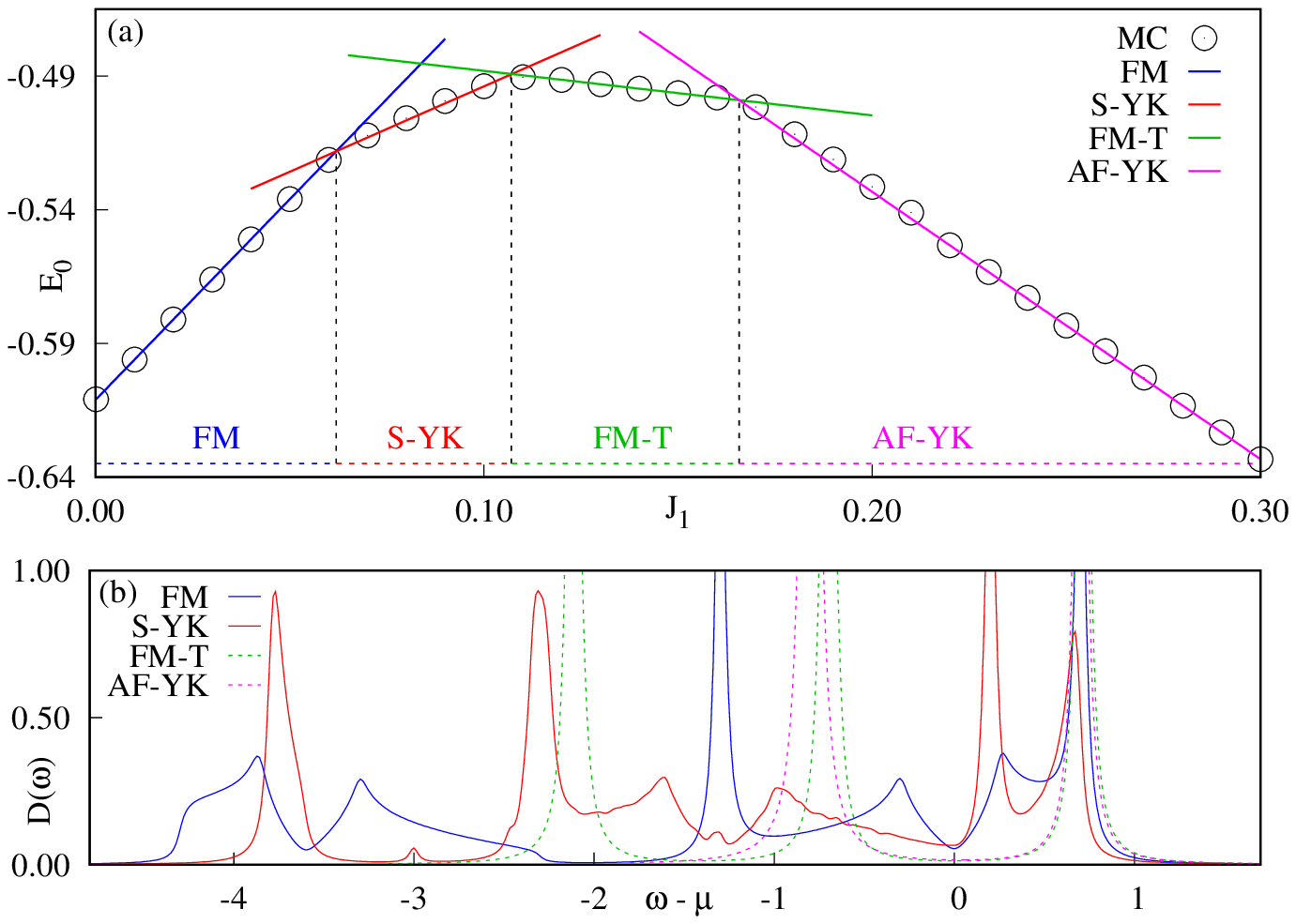}
	\caption{ (Color online) (a) Ground state energy per site $E_0$ vs $J_1$ plot at $n_e=2/3$ obtained by hMCMC simulation (circles) at temperature $T=0.001t_1$. Straight lines are the energies of different phases as indicated. (b) Electron density of states corresponding to phases as mentioned.}
	\label{fig:Phases for ne 2/3}
\end{figure}

We run hMCMC simulations on $N=56$ site DHL system at $n_e=2/3$ with varying $J_1$. The extracted lowest temperature energy per site $E_0$ as a function of $J_1$ is depicted by an empty circle in Fig.\ref{fig:Phases for ne 2/3}(a). Notice that, if a localized spin configuration (magnetic phase) is stabilized for a range of $J_1$, the electronic Hamiltonian (a function of localized spin orientation) gives a constant electronic energy over that range of $J_1$. So, the total energy of the system becomes a linear function of the parameter $J_1$ only. So, any linear region in $E_0$ vs $J_1$ plot indicates a stabilized phase. The straight lines in Fig.\ref{fig:Phases for ne 2/3}(a) correspond to the energy obtained for ideal long-range
ordered spin arrangements (further optimized, analytical or putting spin configuration by hand).

\begin{figure}[hbt!]
	\centering
	\includegraphics[width=0.95\linewidth,angle=00,trim={0.0cm 8.5cm 0.0cm 0.0cm},clip]{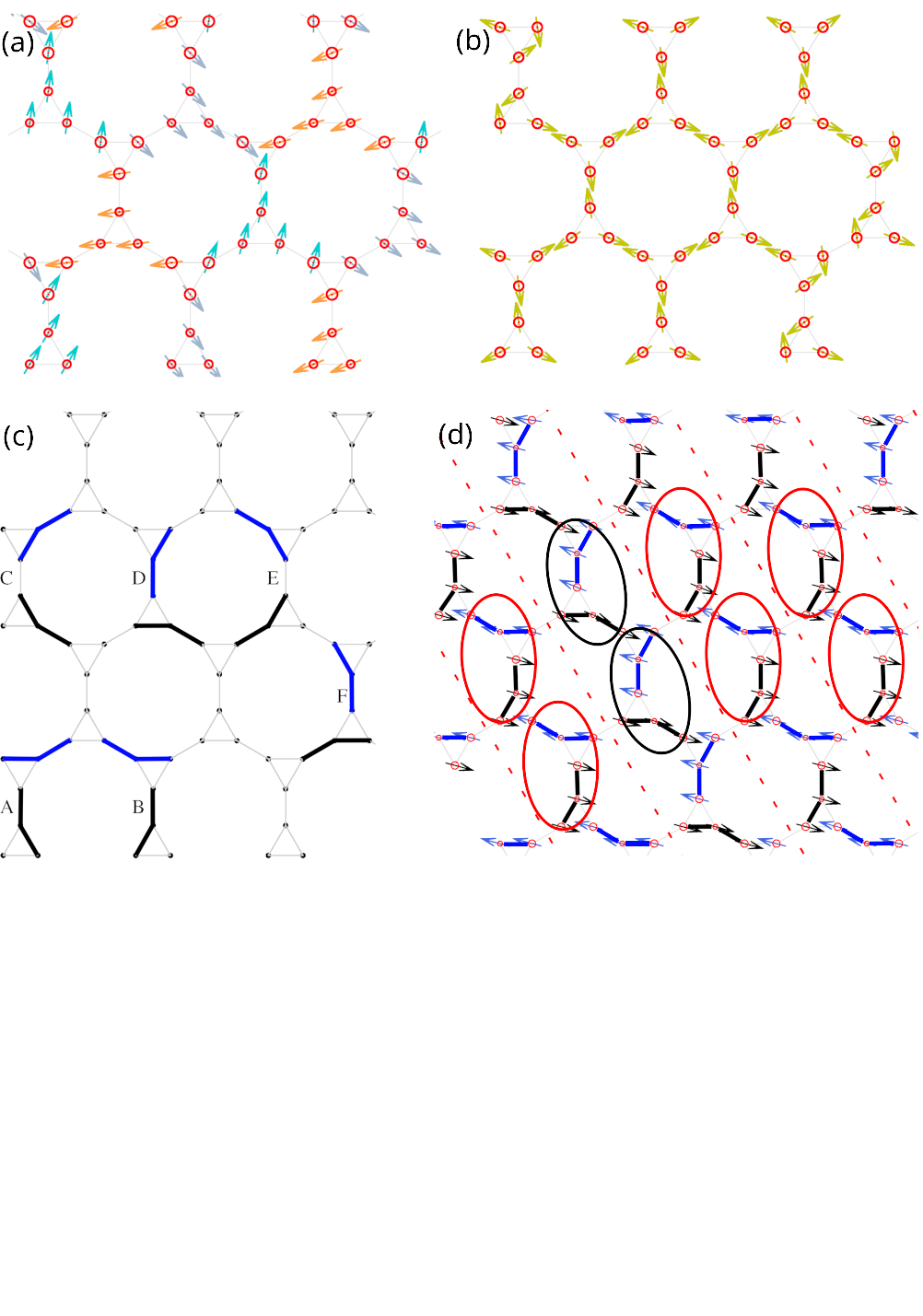}
	\caption{ (Color online) The real space spin configuration (arrows) and electron charge density profile (indicated by size of red circles) obtained from hMCMC simulation at $\text{n}_\text{e} = 2/3$ for (a) S-YK phase at $J_1 = 0.08$ and (b) AF-YK phase at $J_1 = 0.27$. (c) Six Possible configurations (A, B ...F) of two anti-aligned trimers (black and blue segments) within a unit cell to account for the degeneracy of FM-T phase, and (d) The hMCMC snapshot of FM-T phase at $J_1 = 0.14$ represents one of many degenerate configurations tiled with B and D types running along $(\vec{a_2}$-$\vec{a_1})$ lattice direction}
	\label{fig:Final_All_Configuration}
\end{figure}

As shown in Fig.\ref{fig:Phases for ne 2/3}(a), at small $J_1$ the electrons gain kinetic energy by polarizing the localized spins, leading to FM phase with uniform electron charge density $n_i^0$ up to $J_1\sim 0.06$ explained by so-called {\it double exchange} mechanism. On the other limit at large $J_1$, the localized spins on each triangle point $120^{\circ}$ with each other and inter-triangular bonds to be AFM aligned. We label this as antiferromagnetic YK (AF-YK) phase of which the real-space spin configuration obtained from hMCMC simulations is shown in Fig.\ref{fig:Final_All_Configuration}(b). As the AFM bonds prohibit electrons to hop outsite the YK triangles, the energy $E_0$ can be calculated simply by diagonalizing the $3\times 3$ hopping matrix for one YK triangle and fill two electrons in the lowest two eigenstates in addition to counting the relevant FM/AFM bond energies. This leads energy line $E_0/t_1=-1/3-J_1$ which falls perfectly on the hMCMC energies as shown in Fig.\ref{fig:Phases for ne 2/3}(a). This phase also has uniform $n_i^0$ (indicated by the size of red circle at each site in Fig.\ref{fig:Final_All_Configuration}(b)). However, a pair of electrons is confined to hop in each YK triangle, giving an insulating state unlike FM phase which is metallic.

\begin{figure}[hbt!]
	\centering
	\includegraphics[width=0.47\linewidth,angle=-90,trim={6.7cm 0.3cm 0.05cm 3.0cm},clip]{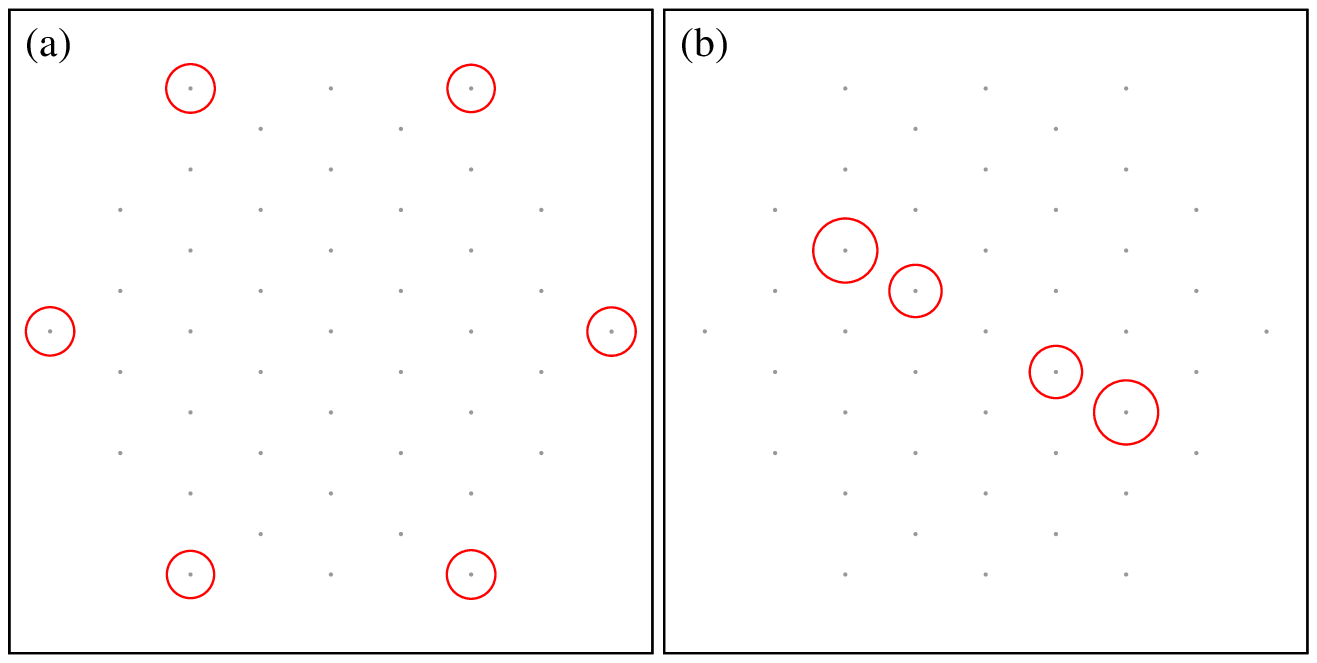}
	\caption{(Color online) Structure factor of (a) S-YK, and (b) S-AF phase.}
	\label{fig:Structure Factor}
\end{figure}

In the intermediate range of $J_1$, the hMCMC energy values are fitted with two straight lines corresponding to two exotic coupled spin-charge ground states. As we increase $J_1$ beyond $0.06$ upto $0.12$, the AFM interaction between localized spins starts competing with FM order mediated by itinerant electrons. This relatively smaller $J_1$ seems to disrupt the FM state, but the electrons still continue to gain sufficient kinetic energy within 6-site FM clusters which are then arranged in a YK $120^{\circ}$ due to $J_1$. We label it as the supercluster YK (S-YK) phase. The hMCMC snapshot of spin orientation and $n_i^0$ represented by the size of red circle at each site are shown in  Fig.\ref{fig:Final_All_Configuration}(a). The S-YK spin texture is also captured by the spin structure factor peak $S(q)$ at the corner of first Brillouin zone as shown in Fig.\ref{fig:Structure Factor}(a). Note that the S-YK phase supports $n_i^0$ modulation (low and high electron density triangles) in the form of a superstructure triangular lattice. Further increase in $J_1$ results even more interesting coupled spin-charge phase consisting of FM 3-site line segments.

Fig.\ref{fig:Final_All_Configuration}(d) depicts FM-T phase spin configuration obtained by hMCMC simulation at $J_1=0.14$ where blue trimers are anti-aligned to black ones. We label this state as ferromagnetic trimer (FM-T) phase. Again, as the electronic hopping integrals $\tilde{t}_{1,2}$ vanish on AFM bonds, the calculation of energy $E_0$ needs $3\times 3$ matrix diagonalization and relevant FM/AFM bond energies. This leads to the energy line as: {$E_0/t_1=-\frac{\sqrt{2}}{3} - \frac{J_1}{6}$ which perfectly falls on the hMCMC energy values confirming the FM-T phase within the range $0.11\lesssim J_1\lesssim 0.16$ as shown in Fig.\ref{fig:Phases for ne 2/3}(a). The density profile $n_i^0$ in FM-T phase shows a larger charge at end sites as compared to the middle one in each trimer, as indicated by the size of red circles at each site. This is again an insulating state as two electrons are trapped in each trimer.

Interestingly, the FM-T phase has macroscopic degeneracy; i.e, there exists a large number of possible ways to tile the lattice with such trimers antiparallel to NN trimers. To calculate degeneracy, we identify six possible ways (A, B...F) to arrange an anti-FM aligned pair of trimers within a unit cell as shown in Fig.\ref{fig:Final_All_Configuration}(c). The ground state configurations can accommodate the stripes running along one of three lattice directions (e.g, $\vec{a_1}$, $\vec{a_2}$, and $(\vec{a_2}$-$\vec{a_1})$}; see Fig.\ref{fig:Unit_Cell}). One such ground state spin arrangement featuring stripes (bounded with dashed lines) along $(\vec{a_2}$-$\vec{a_1})$ lattice direction and made of B-type (enclosed in red ellipses) and D-type (enclosed in black ellipses) is shown in Fig.\ref{fig:Final_All_Configuration}(d). In general, each stripe along $(\vec{a_2}$-$\vec{a_1})$ direction can be occupied with one of three pairs: B/D-type, E/F-type, and A/C-type; resulting $3\times2^{N_s}$ degenerate states, $N_s$ being the number of stripes. Similar analysis along $\vec{a_2}$ directional pairs: C/D-type; A/F-type; B/E-type configuration would also amount to  $(3\times2^{N_s}-6)$ degenerate states. However, along $\vec{a_1}$ direction, the ground state configurations choose only one out of (A, B,..., F) types which have already been counted earlier. Taking trimer spin directions into account, the ground state degeneracy comes out to be $12\times(2^{N_s} - 1)$. 

To further explain the stabilization of these magnetic phases, we calculate the electronic density of states $D(\omega)$ as shown in Fig.\ref{fig:Phases for ne 2/3}(b) for various representative phases. As compared to FM phase, the larger $D(\omega)$ peaks are pushed down to the lower energy side, which helps to gain lower electronic energy and stabilizing the S-YK phase. Even more interestingly, as both AF-T and AF-YK phases consist of electronically disconnected 3-site open chains and YK $120^{\circ}$ triangles, respectively, the $D(\omega)$ of these phases are comprised of filled flat bands separated from the empty ones. These large gaps at the Fermi level make the system insulating and stabilize the phases, which can be seen as a {\it band effect}.

\subsection{Phases at n$_\text{e}$ = 1/2}
\begin{figure}[H]
	\centering
	\includegraphics[width=0.7\linewidth,trim={0.4cm 0.9cm 0.4cm 0.5cm},clip,angle=270]{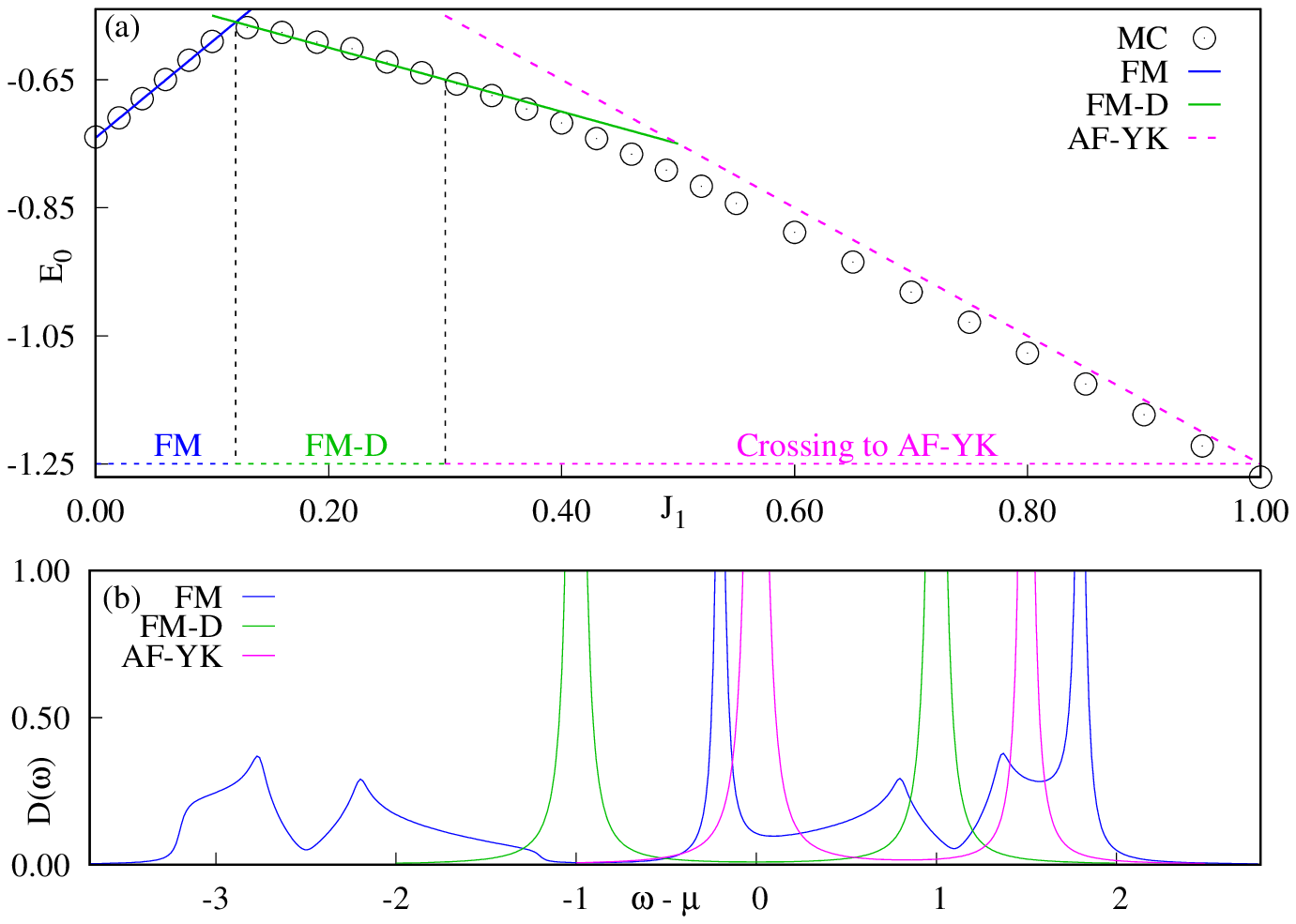}
	\caption{ (Color online) (a) Energy per site Vs $\text{J}_\text{H}/\text{t}_0$ is plotted here for electron concentration $\text{n}_\text{e} = 1/2$ per site. Black circles are energies obtained from hMCMC simulation. The color lines are to indicate the energy of different phases mentioned. (b) Density of states of electron at the ground state for three different phases (FM, FM-D \& AF-YK).}
	\label{fig:Phases for ne 1/2}
\end{figure}

To explore the phases at electronic filling $n_e=1/2$, we run hMCMC simulations and plot the ground state energy $E_0$ for various values of $J_1$ as shown (empty circle) in Fig.\ref{fig:Phases for ne 1/2}(a). As earlier, again we get a linear region representing FM phase for smaller $J_1\lesssim 0.12$ explained by double exchange mechanism. As the value $J_1$ is increased, the system transits to an interesting magnetic phase consisting of ferromagnetic dimers (FM-D) that are anti-aligned to their neighbours. 

\begin{figure}[hbt!]
	\centering
    \includegraphics[width=0.87\linewidth,angle=00,trim={2.1cm 8.5cm 3.9cm 8.5cm},clip]{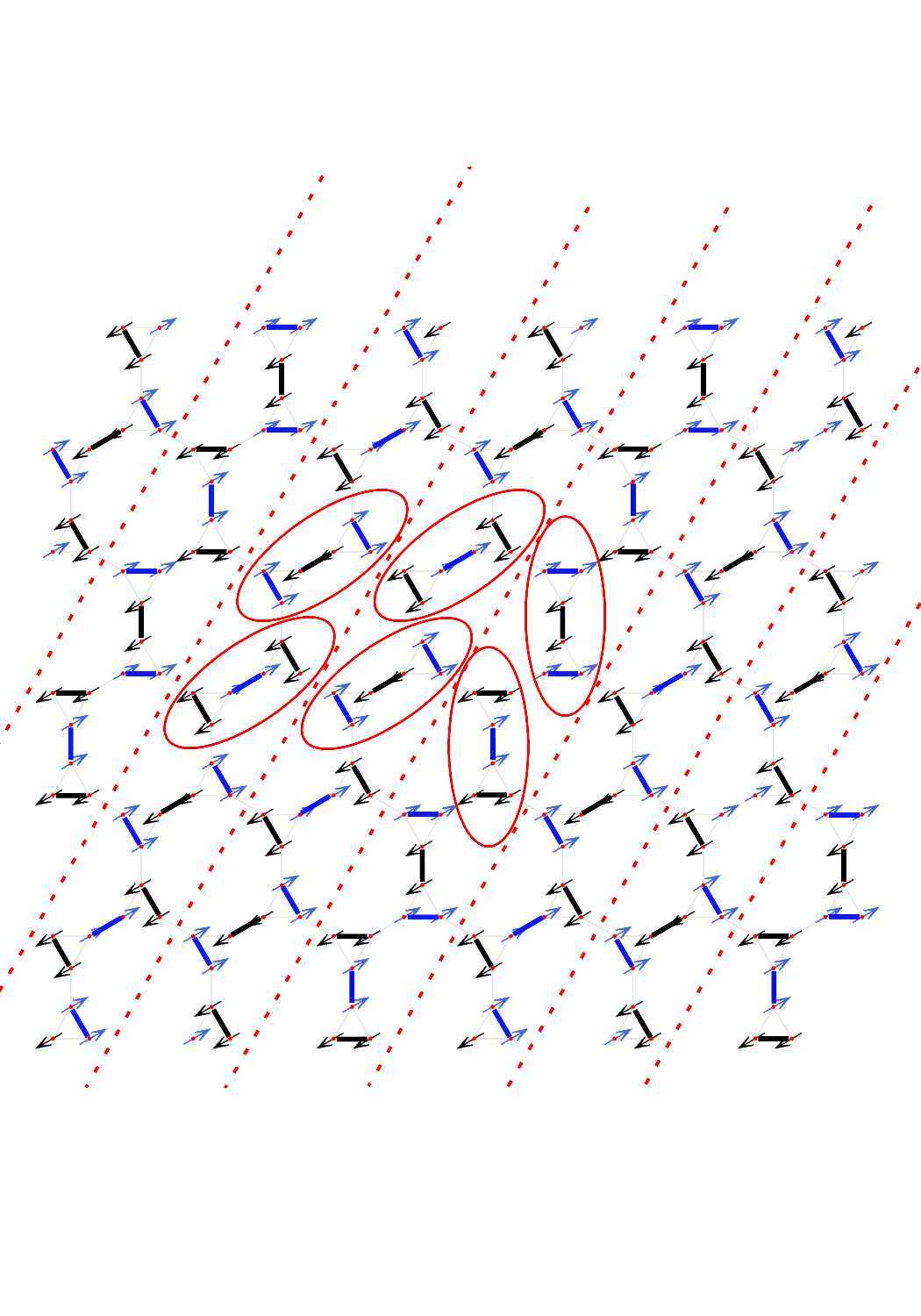}
	\caption{(Color online) Snapshot of spin configuration of FM-D phase obtained from hMCMC simulation for $n_e=1/2$ at $J_1=0.23$ showing the units (ellipses) running along $\vec{a_2}$ lattice direction.}
	\label{fig:1/2 Dimer phase}
\end{figure}

A snapshot of real-space spin configuration obtained by hMCMC simulation for $J_1=0.23$ is depicted in Fig.\ref{fig:1/2 Dimer phase} (black and blue arrows). As one electron is trapped within each dimer, the hopping between two sites creates bonding and anti-bonding states with energy $\pm 1$. This is also reflected by two flat band peaks in $D(\omega)$ plot at energies $\pm 1$ as shown (green line) in Fig.\ref{fig:Phases for ne 1/2}(b). One electron in each dimer occupies the bonding state, resulting in an insulating state. The energy $E_0$ for the ideal FM-D phase can be obtained as: $E_0/t_1=-\frac{1}{2}-\frac{J_1}{2}$ (green straight line), which fits very well with hMCMC energies.

Like the FM-T phase, this state is also macroscopically degenerate, as there are multiple ways to tile the whole lattice with anti-FM aligned dimers. Just like FM-T phase, the units containing three dimers (enclosed by red ellipse as shown in Fig.~\ref{fig:1/2 Dimer phase}) can be arranged in one of three lattice directions: $\vec{a_1}$, $\vec{a_2}$, and $(\vec{a_2}$-$\vec{a_1})$ along the stripes. Each stripe can again accommodate either vertical or inclined units (red ellipses) as depicted in Fig.~\ref{fig:1/2 Dimer phase}). Also, there exists another type of similar inclined unit oriented along the direction $(\vec{a_2}$-$\vec{a_1})$. Counting the two possible spin directions the degeneracy is given by $6\times (2^{N_s}-1)$, $N_s$ being the number of stripes in each direction. A similar type of macroscopic degeneracy of dimer phase on a simple honeycomb lattice has been discussed in Ref.~\cite{vendor_sanjeev_PhysRevLett.107.076405}.  Further increase in $J_1$ allows the system to cross over to AF-YK phase stable at large $J_1$.

\subsection{Phases at n$_\text{e}$ = 1/3}
\begin{figure}[hbt!]
	\centering
	\includegraphics[width=0.7\linewidth,trim={0.58cm 1.1cm 0.4cm 0.5cm},clip,angle=270]{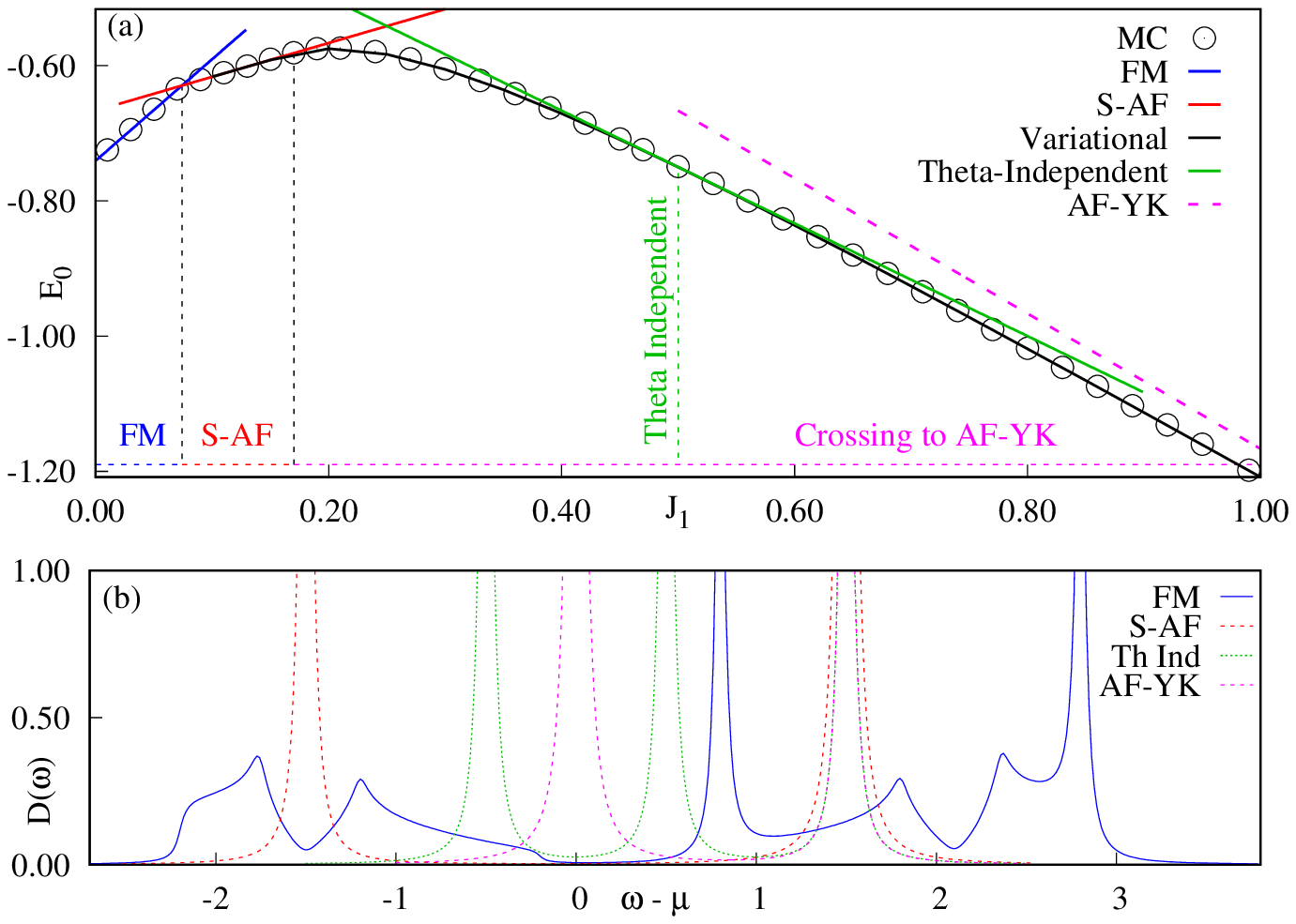}
	\caption{ (Color online) (a) Ground state energy per site $E_0$ vs $J_1$ for $n_e=1/3$ at $T=0.001t_1$ obtained from hMCMC simulations. The straight lines indicate the energies of corresponding phases. The black curved line is obtained by variational calculations (see text) (b) Density of state of different phases.}
	\label{fig:Phases for ne 1/3}
\end{figure}

\begin{figure}[hbt!]
	\centering
	\includegraphics[width=0.7\linewidth,angle=270,trim={8.8cm 9.65cm 1.8cm 7.55cm},clip]{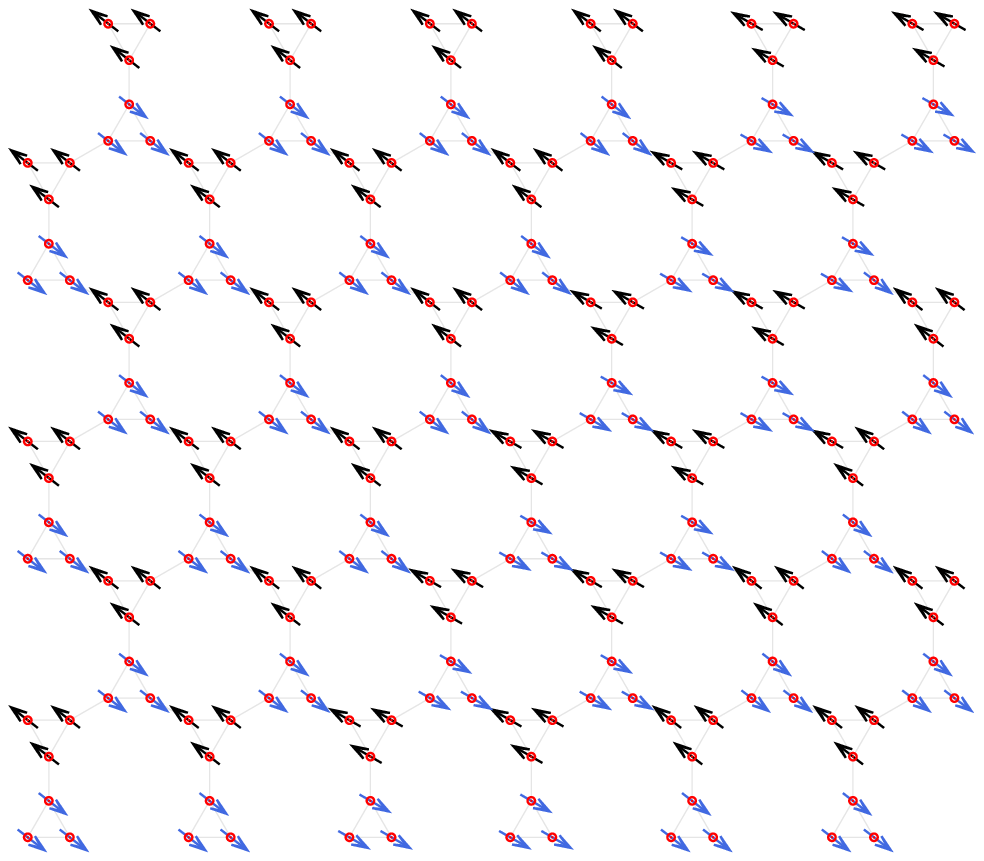}
	\caption{(Color online) Real space spin configuration of S-AF phase for $n_e=1/3$ at $J_1 = 0.12$ obtained from hMCMC simulation. The size of red circles indicates $n_i^0$}
	\label{fig:1/3 Super AFM Phase}
\end{figure}

At electronic filling $n_e=1/3$, the plot for hMCMC energies (empty circles) as a function of $J_1$ is shown in Fig.\ref{fig:Phases for ne 1/3}(a). As expected, the double exchange mechanism forces the system to be in FM phase for low $J_1\lesssim 0.07$. Further increase in $J_1$ destroys the FM spin arrangement. However, the electrons still manage to keep the FM configuration within each triangle, which are anti-aligned to their nearest FM triangles, manifested as a super-AF (S-AF) in the form of a honeycomb lattice structure in the parameter range $0.07\lesssim J_1\lesssim 0.18$. The hMCMC spin configuration and corresponding spin structure factor of S-AF phase are depicted in Fig.\ref{fig:1/3 Super AFM Phase} and Fig.\ref{fig:Structure Factor}(b) respectively. This phase has uniform charge distribution; however, one electron is trapped within each triangle, leading to an insulating state, which is also confirmed by the flat bands with large gaps in $D(\omega)$ plot for S-AF phase.     

When $J_1$ is increased further, the hMCMC energies start to deviate from that of the ideal S-AF spin configuration. Looking at the hMCMC ground state spin arrangements beyond the S-AF phase, we notice that the AF alignment on all inter-triangular bonds, which is the case in the S-AF phase too, always persists. This suggests that the ground state spin configurations beyond FM phase actually consist of electronically disconnected triangular units with one electron trapped within, making the system insulating.   

\begin{figure}[hbt!]
	\centering
	\includegraphics[width=0.65\linewidth,angle=270,trim={0.6cm 0.2cm 2.1cm 3.7cm},clip]{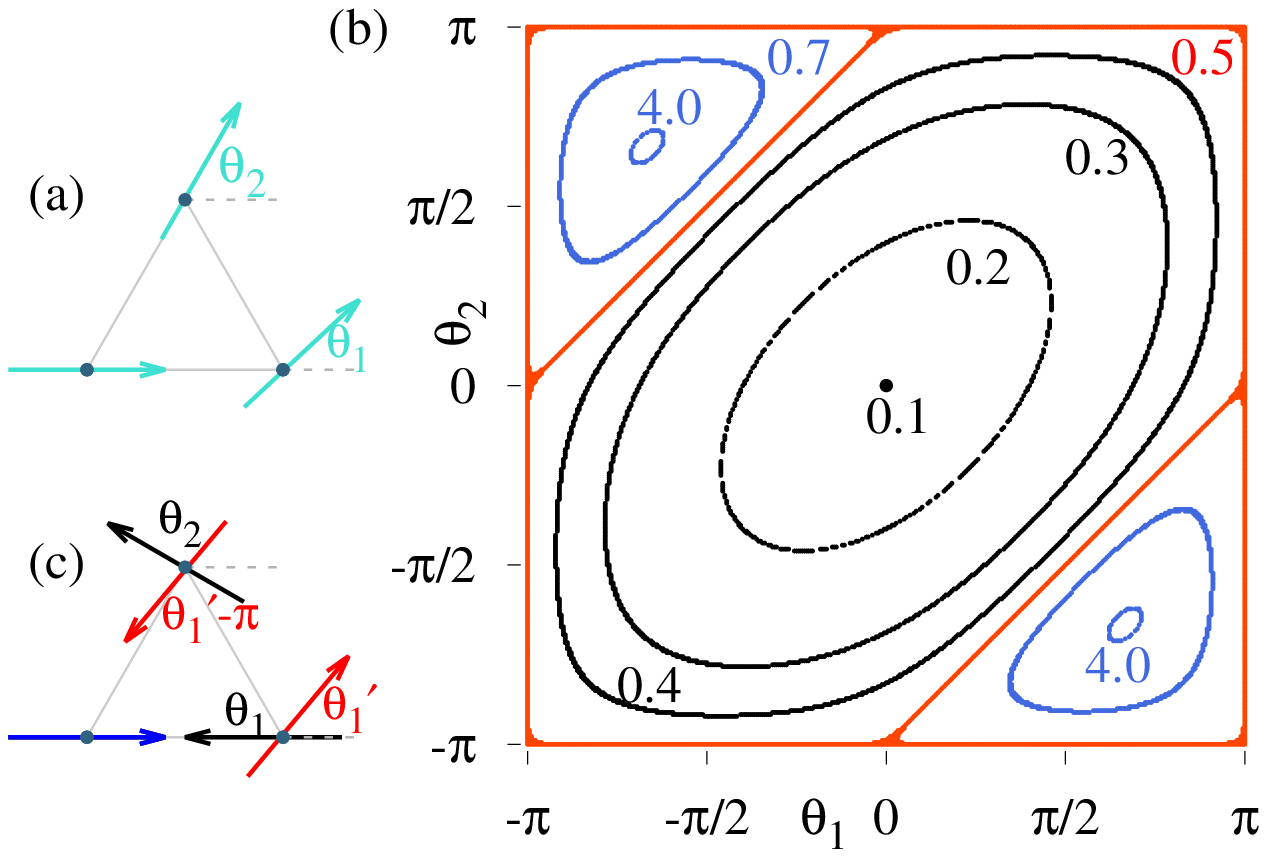}
	\caption{(Color online) (a) Showing the spin arrangement considered with two parameters ($\theta_1, \theta_2$) (b) The contours for ground state energy in $\theta_1-\theta_2$ plane at various $J_1$ labelled at each contour. At $J_1=0.50$, the contour is represented by $\theta_1 = \pm\pi, \theta_2 = \pm\pi$ \& $|\theta_1-\theta_2|=\pi$. (c) Two types of spin arrangement at $J_1=0.5$. represented by red and black arrows (blue spin is fixed in horizontal direction)} 
    \label{fig:1/3 Crossover}
\end{figure}

To substantiate the hMCMC results, we perform simple variational calculations on a triangle to obtain the minimum energies by varying the spin orientations. As depicted in Fig.\ref{fig:1/3 Crossover}(a), one spin direction is fixed, say along the horizontal axis and the other two spin directions (denoted by $\theta_1$ and $\theta_2$) are varied from $-\pi$ to $\pi$. For each pair ($\theta_1$, $\theta_2$), we calculate the electronic energy by diagonalizing $3\times 3$ hopping matrix and filling the lowest eigen energy (one electron per triangle for $n_e=1/3$). The magnetic energy per site can simply be equal to $\frac{J_1}{3}[cos(\theta_1)+cos(\theta_2)+cos(\theta_1-\theta_2)]-1/2$. For each $J_1$ in the range of $J_1\ge 0.07$, we keep track of the minimum energy $E_0$ and the pair ($\theta_1$, $\theta_2$) that minimizes the energy.


As depicted (black solid line labeled as variational) in Fig.\ref{fig:Phases for ne 1/3}(a), the minimum energy $E_0$ obtained from these simple variational calculations perfectly matches with hMCMC energies. We also analyzed the pair ($\theta_1,\theta_2$) at which the energy is minimized to understand how the ground-state spin configurations evolve from the FM state at small $J_1$ to the AF-YK phase at large $J_1$. Fig.\ref{fig:1/3 Crossover}(b) depicts the minimum energy contours in $\theta_1-\theta_2$ plane for different $J_1$ values as indicated. As shown, the centre point ($\theta_1=0, \theta_2=0$) represents S-AF state where all spins in a triangle are FM aligned. As $J_1$ is increased, the system enters into a highly degenerate ground state, of which the energy follows almost the elliptical contour with major axis along $\theta_1=\theta_2$ line. An interesting spin configuration is reached around $J_1\sim 0.5$. The energy contour (red line) indicates that if $\theta_1 (\theta_2)-$spin is AFM aligned with the fixed one, any orientation of $\theta_2 (\theta_1)$ is possible because the electronic and magnetic energy becomes independent of $\theta_2 (\theta_1)$, see black arrow configuration in Fig.\ref{fig:1/3 Crossover}(c). Due to symmetry, any value of $\theta_1$ with $|\theta_1-\theta_2|=\pi$ is also a possible ground state configuration, see red arrow in Fig.\ref{fig:1/3 Crossover}(c). The overall spin orientation has all inter-triangular bonds and one of three intra-triangle bonds in each triangle to be anti-FM aligned. The corresponding energy with this particular spin configuration is depicted by a green straight line in Fig.\ref{fig:Phases for ne 1/3}(a), which is almost degenerate with hMCMC energies in a finite $J_1$ range around $J_1=0.50$.  Further increase in $J_1$ slowly pushes the system to the point $\theta_1=\pm 2\pi/3, \theta_2=\mp 2\pi/3$ i.e, AF-YK phase at large $J_1$. 

\section{Conclusions}
In summary, we have identified four distinct coupled spin–charge ordered ground states in the strong-coupling regime of the Kondo lattice model on a frustrated decorated honeycomb lattice. Using a hybrid Markov chain Monte Carlo approach, which integrates classical sampling of localized spins with exact diagonalization of the itinerant electron Hamiltonian, we characterized the emergence of super-antiferromagnetic (S-AF) and super-Yafet–Kittel (S-YK) phases composed of three-site and six-site ferromagnetic clusters, respectively. Additionally, we uncovered two unconventional cluster-based states—FM dimers (FM-D) and FM trimers (FM-T)—each exhibiting antiferromagnetic alignment between neighbouring building blocks and hosting a macroscopic degeneracy associated with their configurational freedom. Except for the S-YK phase, these spin textures confine electrons to locally fragmented clusters, producing flat electronic bands through substantial spectral gap formation, thereby stabilizing the corresponding ordered states via a band-structure mechanism. Our findings reveal an unexpected richness of emergent cluster magnetism and correlated flat-band physics, with potential relevance to electron-doped frustrated magnets and metal–organic framework based quantum materials.

{\it Acknowledgements.---}
A part of the computation is performed in Physics Department, Jadavpur University Computer Facility (Sanction No. SR/FST/PS-1/2022/219(C)) and HPC workstation procured from SERB-DST project (reference SRG/2020/001203).
\appendix

\bibliography{decorated_HC}

\end{document}